\documentclass[conference]{IEEEtran}
\IEEEoverridecommandlockouts
\usepackage{cite}
\usepackage{amsmath,amssymb,amsfonts}
\usepackage{algorithm}
\usepackage{graphicx}
\usepackage{textcomp}
\usepackage{hyperref}
\usepackage{booktabs}
\usepackage{amsmath}
\usepackage{xcolor}
\usepackage{algorithm}
\usepackage{authblk}
\usepackage{etoolbox}
\usepackage{marvosym}
\usepackage{algpseudocode}

\usepackage[utf8]{inputenc}
\usepackage{tcolorbox}

\def\BibTeX{{\rm B\kern-.05em{\sc i\kern-.025em b}\kern-.08em
    T\kern-.1667em\lower.7ex\hbox{E}\kern-.125emX}}
\usepackage{algpseudocode}
    
\begin{document}

\title{RA-Gen: A Controllable Code \textbf{Gen}eration Framework Using \textbf{R}e\textbf{A}ct for Multi-Agent Task Execution
}

\author[1,2]{Aofan Liu}
\author[3]{Haoxuan Li} \author[1,2]{Bin Wang}
\author[1,2]{Ao Yang}
\author[1,2]{Hui Li* \Letter}
\affil[1]{\footnotesize School of Electronic and Computer Engineering, Peking University}
\affil[2]{\footnotesize Guangdong Provincial Key Laboratory of Ultra High Definition Immersive Media Technology, Shenzhen Graduate School, Peking University}
\affil[3]{\footnotesize Shenzhen International Graduate School, Tsinghua University}

\maketitle

\begin{abstract}

Code generation models based on large language models (LLMs) have gained wide adoption, but challenges remain in ensuring safety, accuracy, and controllability, especially for complex tasks. Existing methods often lack dynamic integration of external tools, transparent reasoning, and user control over safety.
To address these issues, we propose a controllable code generation framework utilizing the ReAct paradigm for multi-agent task execution. This framework is a multi-agent system designed to enable efficient, precise, and interpretable code generation through dynamic interactions between LLMs and external resources. The framework adopts a collaborative architecture comprising four specialized agents: a Planner for task decomposition, a Searcher that leverages the ReAct framework for reasoning and tool integration, a CodeGen agent for accurate code generation, and an Extractor for structured data retrieval. The ReAct-based Searcher alternates between generating reasoning traces and executing actions, facilitating seamless integration of internal knowledge with external tools (such as search engines) to enhance accuracy and user control. 
Experimental results show the framework’s effectiveness across multiple languages, achieving a 94.8\% security rate on the SVEN dataset with CodeQL, outperforming existing approaches. Its transparent reasoning process fosters user trust and improves controllability.

\end{abstract}

\begin{IEEEkeywords}
Automated Programming, Code Generation, Multi-agent, ReAct Framework, Reasoning Trajectory, Large Language Model
\end{IEEEkeywords}

\section{Introduction}

The rapid advancement of artificial intelligence has significantly enhanced the capabilities of large language models (LLMs) in the domain of code generation. Despite their widespread adoption, ensuring the safety of generated code remains a paramount challenge. Traditional code generation approaches often rely solely on the inherent capabilities of a single model, lacking mechanisms for comprehensive safety assurance. This limitation becomes particularly pronounced when addressing complex programming tasks, where the generated code must not only be functional but also secure and free from vulnerabilities.

Existing retrieval-based code generation methods generally rely on a fixed retrieval process \cite{guo2024retrieval}, where external tools (such as search engines or knowledge bases) are first used to retrieve relevant information, followed by the generation of code. While effective in many instances, these methods lack the flexibility to dynamically switch between retrieval and reasoning processes. Consequently, the model cannot adjust its retrieval strategy based on changes in its reasoning state. When the model encounters novel problems or requires further clarification, it may need to retrieve additional information. However, current approaches do not facilitate the dynamic adjustment of retrieval strategies in response to evolving reasoning requirements.

Furthermore, while retrieval and reasoning in existing methods can improve the efficiency and accuracy of code generation to some extent, these methods typically rely on a static process and lack the flexibility and adaptability necessary for complex, multi-faceted tasks. Code generation, especially in complex contexts, often involves multiple subtasks that demand diverse skills and specialized knowledge. A single agent, operating within the constraints of a fixed process, often struggles to effectively handle such complexities.

To overcome these challenges, this paper introduces a multi-agent controllable code generation system grounded in the ReAct framework \cite{yaoReActSynergizingReasoning2023}. This system is designed to achieve efficient, precise, and safe code generation and task execution through the dynamic interaction between LLMs and external tools. The proposed framework employs a collaborative architecture consisting of four specialized agents: a \textbf{Planner} for task decomposition, a \textbf{Searcher} that leverages the \textbf{ReAct} framework\cite{yaoReActSynergizingReasoning2023} for reasoning and tool integration, \textbf{CodeGen} for generating accurate code, and an \textbf{Extractor} for structured data retrieval.

The \textbf{Searcher} agent, based on the \textbf{ReAct} framework, alternates between generating reasoning trajectories and executing actions. This approach facilitates the seamless integration of internal knowledge with external tools, such as security scanners and search engines, thereby enhancing both the safety and controllability of the code generation process. By dynamically adjusting reasoning paths and action strategies, the \textbf{Searcher} agent provides the framework with significant flexibility, enabling it to retrieve external resources as needed to supplement its reasoning and adapt to varying task requirements.

Furthermore, the system explicitly records \textbf{reasoning trajectories}, ensuring that each decision step is traceable and transparent. This transparency not only improves the interpretability of the generated code but also fosters greater user trust in the system's outputs. The collaborative effort of the four agents automates the entire process from code understanding to generation, effectively handling complex tasks that require multi-step execution and supporting multiple programming languages, including Python and C/Cpp.

The contributions are threefold:

\begin{enumerate}
    \item \textbf{ReAct-Based Multi-Agent Collaborative Architecture}: We propose a novel multi-agent architecture that integrates the ReAct framework, comprising a Planner, Searcher, CodeGen, and Extractor. This architecture automates the end-to-end process of code generation through effective task decomposition and inter-agent collaboration, enhancing both the flexibility and efficiency of the system.
    \item \textbf{Dynamic Reasoning Mechanism with External Tool Integration}: The Searcher agent utilizes the ReAct framework to alternate between generating reasoning trajectories and performing search actions. This dynamic mechanism enables the seamless combination of external tools, such as security scanners and search engines, with the model's internal knowledge. By alternating between reasoning and search, it significantly enhances the safety, controllability, and explainability of the generated code.
    \item \textbf{Comprehensive Experimental Validation}: Our experiments demonstrate that the proposed framework efficiently handles code generation tasks across multiple programming languages and excels in complex scenarios requiring multi-step task execution. Achieving a 94.8\% security rate using CodeQL on the SVEN dataset, the system surpasses existing methods. Additionally, the explicit documentation of reasoning trajectories provides a transparent decision-making process, offering new insights for the interpretability and safety of code generation tasks.
\end{enumerate}

In summary, the proposed multi-agent controllable code generation framework addresses the critical challenge of ensuring the safety of generated code by leveraging the strengths of the ReAct framework and multi-agent collaboration. This approach not only enhances the quality and reliability of the generated code but also paves the way for more trustworthy and secure AI-driven code generation systems.

\section{Related Work}

\begin{figure*}[htbp]
\centering
\includegraphics[width=0.70\linewidth,keepaspectratio]{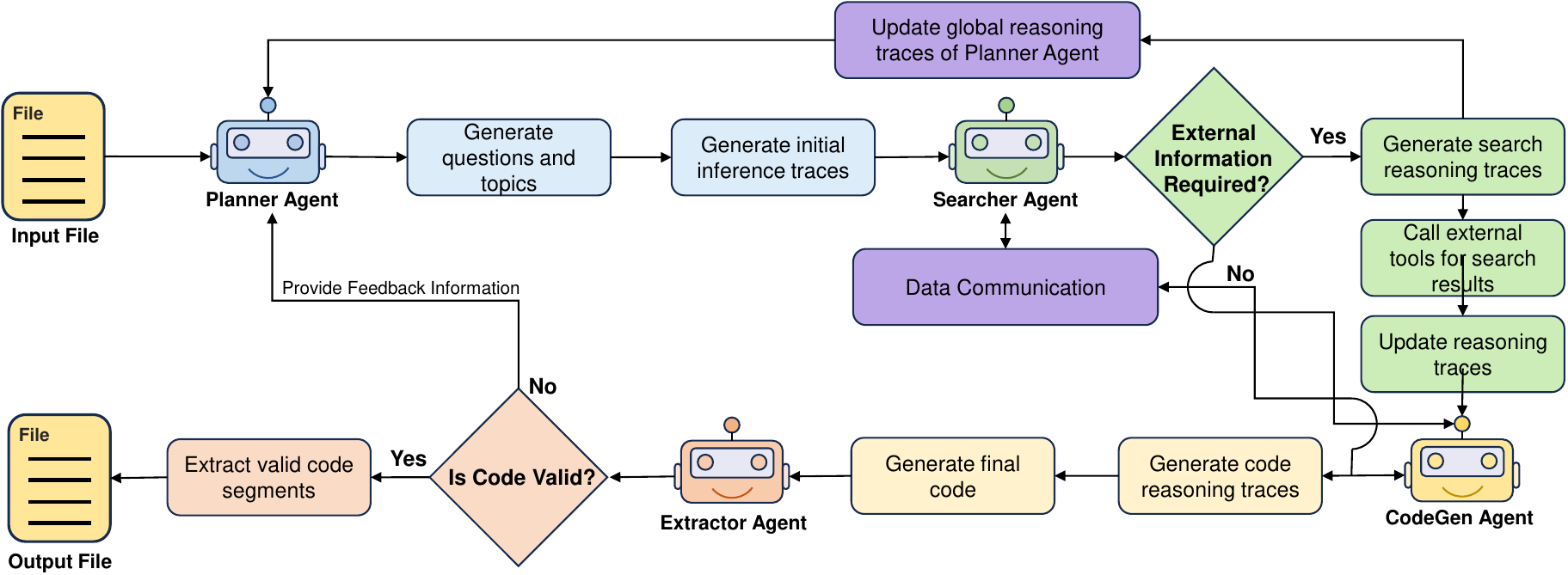}
\caption{Architecture of the multi-agent framework for secure code generation. The framework comprises four key components: the \textbf{Planner}, which decomposes tasks and generates initial reasoning trajectories; the \textbf{Searcher}, which refines trajectories by combining reasoning and external tools; the \textbf{CodeGen}, which generates secure code patches; and the \textbf{Extractor}, which validates and extracts functional code snippets. This collaborative process ensures the generation of high-quality, secure code.}
\label{fig}
\end{figure*}

\subsection{Code Generation}

The application of large language models (LLMs) to code-related tasks has garnered significant attention, driven by the availability of open-source codebases and the increasing demand for tools to boost developer productivity. LLMs have demonstrated exceptional performance in areas such as code generation \cite{talridnikCodeGenerationAlphaCodium, chenEvaluatingLargeLanguage2021}, program repair \cite{keRepairingProgramsSemantic2015, olaussonSelfRepairSilverBullet2023}, automated testing \cite{shiCoCoASTRepresentingSource2023}, code translation \cite{lachauxUnsupervisedTranslationProgramming2020}, and code summarization \cite{ahmedFewshotTrainingLLMs2023}, highlighting their versatility across the software development lifecycle.

Models like CODEX \cite{OpenAICodexOpenAI}, CodeGen \cite{nijkampCodeGenOpenLarge2023}, InCoder \cite{fried_incoder_2022}, and PolyCoder \cite{xuSystematicEvaluationLarge2022} are specifically designed for code generation, excelling at translating natural language descriptions into functional code. These models reduce cognitive load and accelerate development by bridging natural language with programming languages \cite{CodeRAGBenchCanRetrieval, CompetitionlevelCodeGeneration}, making them invaluable in modern software development.

\subsection{Multi-Agent Systems}

Multi-agent systems are engineered to autonomously manage complex tasks, leveraging their strengths in decision-making, tool utilization, and memory management. LLM-based agents, guided by prompts, can decompose intricate tasks into manageable subgoals, explore multiple solution pathways, and refine their decisions through experiential learning \cite{khot2023decomposedpromptingmodularapproach, NEURIPS2023_271db992, NEURIPS2023_1b44b878}. This capability enhances their autonomy and effectiveness in problem-solving scenarios. Additionally, these agents can integrate external tools, such as APIs and databases, to extend their functionality and adapt to diverse environments \cite{li2023apibankcomprehensivebenchmarktoolaugmented, gao2024retrievalaugmentedgenerationlargelanguage}. Their memory capabilities further bolster performance, enabling the retention and application of information over time through short-term or long-term memory structures \cite{dong2024surveyincontextlearning, NEURIPS2020_6b493230, wang2024survey}.

\subsection{Controlled Code Generation}

Early code generation methods relied on template matching or manually defined rules, which were effective in specific cases but struggled with the complexity and variability of modern tasks. Template-based methods were limited to fixed structures \cite{liuTBarRevisitingTemplatebased2019}, while rule-based approaches required extensive manual effort, making them impractical for large-scale projects \cite{Yang2020A}. With the rise of information retrieval techniques, retrieval-based methods have gained popularity \cite{He_2023}. These methods generate code by retrieving relevant snippets from code repositories, such as GitHub, and synthesizing them into final code \cite{wang2024aigeneratedcodereallysafe}. To improve quality and security, post-processing techniques like code formatting, comment insertion, and variable name optimization have been introduced \cite{nunez2024autosafecodermultiagentframeworksecuring}, alongside the use of static code analysis tools to ensure adherence to coding standards \cite{NEURIPS2021_9e1a3651}.

\subsection{Integration of Multi-Agent Systems and Controlled Code Generation}

The intersection of multi-agent systems and controlled code generation represents a promising avenue for enhancing the safety, accuracy, and controllability of generated code. By leveraging the collaborative strengths of multiple agents, systems can dynamically integrate external tools and apply comprehensive safety checks throughout the code generation process\cite{rasheed2024codepori}. This integration facilitates multi-level reasoning and ensures that the generated code meets stringent security and reliability standards.

Recent research has begun to explore this synergy, demonstrating that multi-agent architectures can effectively manage complex coding tasks while maintaining high standards of code quality and security\cite{huang2023agentcoder}. These systems benefit from the collective intelligence of specialized agents, each contributing unique capabilities that collectively enhance the overall performance and reliability of code generation processes.

\section{Methodology}

We present a Controllable Code Generation Framework utilizing the ReAct framework for multi-agent task execution. This framework integrates a multi-agent architecture designed to facilitate efficient, precise, and controllable code generation through dynamic interactions between large language models (LLMs) and external tools. The system comprises four key agents: the \textbf{Planner}, the \textbf{Searcher}, the \textbf{CodeGen}, and the \textbf{Extractor}. Each agent is responsible for specific functions within the code generation pipeline, collaborating through task decomposition and reasoning traces to effectively manage complex code generation tasks, as shown in Figure \ref{fig}.

\begin{figure*}[htbp]
\centering
\includegraphics[width=\linewidth]{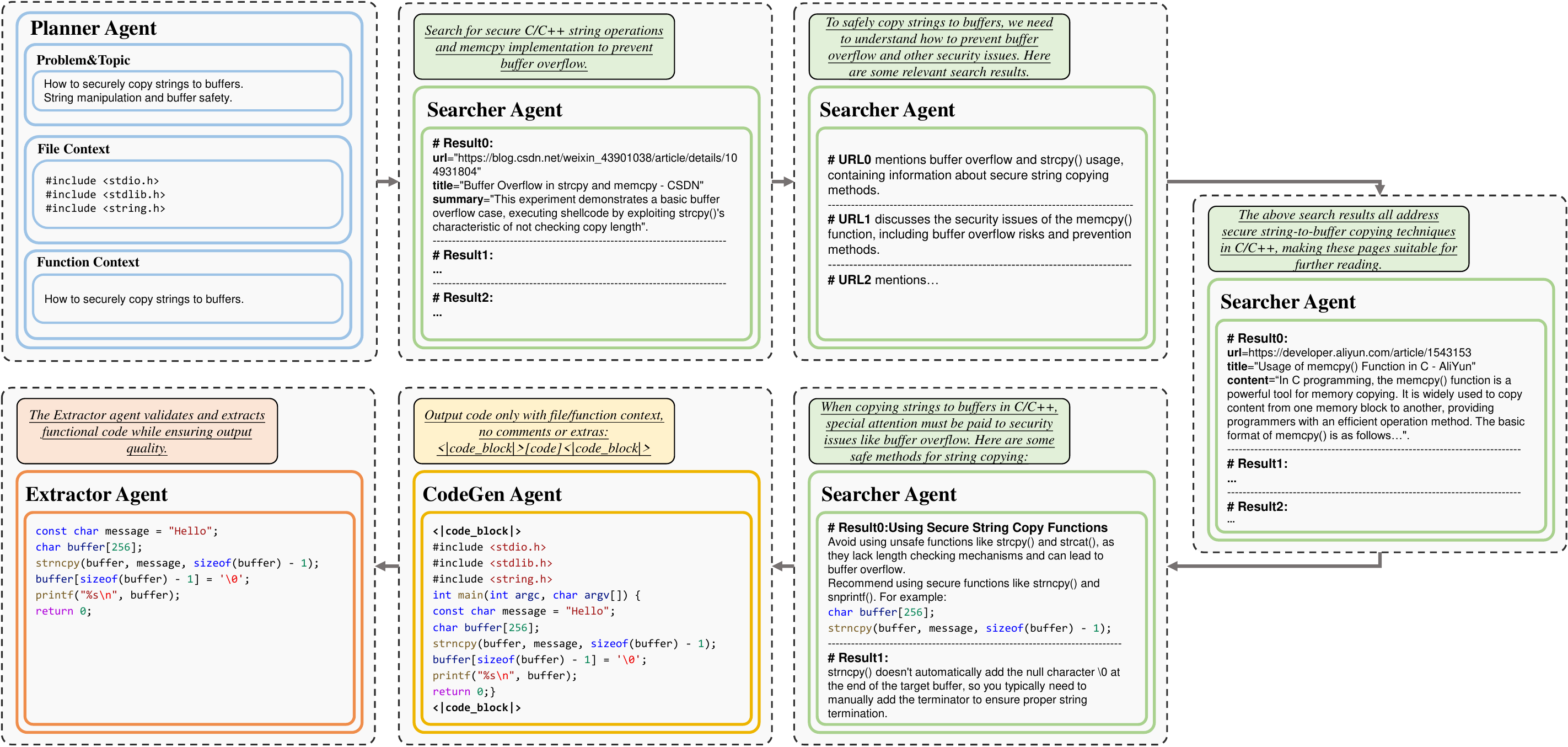}
\caption{Case example illustrating secure string copying to buffers in C/Cpp to prevent buffer overflow with tool "Online Search". The process involves RA-Gen agents: the \textbf{Planner} identifies the problem of string manipulation and buffer safety, the \textbf{Searcher} retrieves relevant information on secure string operations, the \textbf{Extractor} validates and extracts functional code, and the \textbf{CodeGen} agent generates safe implementation examples.}
\label{fig:architecture}
\end{figure*}

\subsection{Multi-Agent Collaborative Architecture}

Let task specification be represented as a tuple \( T = (\mathcal{I}, \mathcal{O}) \) where \( \mathcal{I} \subseteq \mathbb{R}^{d_{\text{in}}} \) denotes input requirements and \( \mathcal{O} \subseteq \mathbb{R}^{d_{\text{out}}} \) specifies desired outputs. Our agent ensemble \( \mathcal{A} = \{ \pi_{\text{Planner}}, \pi_{\text{Searcher}}, \pi_{\text{CodeGen}}, \pi_{\text{Extractor}} \} \) operates through coordinated message passing in state space \( \mathbb{S} \). The collaborative dynamics are governed by:

\begin{equation}
\forall t \geq 0: \quad s_{t+1} = \Phi(s_t, a_t), \quad a_t \sim \bigotimes_{\pi \in \mathcal{A}} \pi(\cdot|s_t)
\end{equation}
where \( \Phi: \mathbb{S} \times \mathcal{A} \rightarrow \mathbb{S} \) is the state transition function and \( \bigotimes \) denotes policy composition.

\textbf{Planner}: Implements hierarchical task decomposition through projection operators:
\begin{align}
\mathcal{D}(T) &= \bigcup_{k=1}^K P_{\mathcal{M}_k}(T) \\
\text{where} \quad P_{\mathcal{M}_k}(T) &= \left\{ s_i \in \mathbb{S} \mid \frac{\partial \mathcal{M}_k}{\partial T} \succeq \xi \right\}
\end{align}
with \( \mathcal{M}_k \in \{\text{sequential}, \text{parallel}, \text{hierarchical}\} \) representing decomposition modalities.

\textbf{Searcher}: The Searcher agent operates by alternating between reasoning and action. It generates reasoning traces \( R = \{ r_1, r_2, \dots, r_m \} \) and executes actions \( A = \{ a_1, a_2, \dots, a_k \} \). Each reasoning trace \( r_i \) is developed step-by-step based on the current task \( T \), while the corresponding action \( a_j \) is determined by the current reasoning step and interactions with external tools. The functions governing reasoning and action are defined as:

\begin{align}
R_i &= f_{\text{reason}}(T, R_{i-1}) \\
A_j &= f_{\text{action}}(R_i, \mathcal{T}_{\text{external}})
\end{align}

where \( f_{\text{reason}} \) represents the reasoning function and \( f_{\text{action}} \) denotes the action function. External tools \( \mathcal{T}_\text{external} \), such as search engines or APIs, provide real-time information to support decision-making.

\textbf{CodeGen}: The CodeGen agent generates code based on subtasks \( S \) and reasoning traces \( R \) from the Planner and Searcher agents. The process uses task decomposition and reasoning to produce accurate code:

\begin{equation}
C = \text{CodeGen}(S, R)
\end{equation}

where \( C \) is the generated code, informed by task specifications and reasoning traces.

\textbf{Extractor}: The Extractor agent analyzes the generated code \( C \), extracting valuable knowledge \( K \) to refine future generations. This knowledge adjusts reasoning traces \( R \) and informs task decomposition \( \mathcal{D}(T) \) for subsequent iterations.

\subsection{Task Decomposition and Dynamic Interactions}

The framework utilizes task decomposition to effectively manage complex code generation tasks. The high-level task \( T \) is recursively broken down into smaller, more manageable subtasks \( S = \{ s_1, s_2, \dots, s_n \} \). Each subtask \( s_i \) is assigned to an agent \( a_i \in \mathcal{A} \) for execution. Agent collaboration is modeled by a timed coordination graph \( G = (V, E, \tau) \), where:

\begin{equation}
\begin{aligned}
V &= \{ v_i \mid v_i = (\pi_i, s_i^{(t)}) \}_{i=1}^4 \\
E &= \{ e_{ij} \mid \tau_{ij} < \delta_{\text{max}} \} \\
\tau_{ij} &\sim \text{Gamma}(k=2, \theta=0.5)
\end{aligned}
\end{equation}

In this graph, \( V \) represents the set of nodes, where each node \( v_i \) corresponds to an agent's policy \( \pi_i \) and its state \( s_i^{(t)} \) at time \( t \). The edges in \( E \) connect the agents, and \( \tau_{ij} \) represents the time delay between agent \( i \) and agent \( j \), constrained by a maximum time \( \delta_{\text{max}} \). The inter-agent communication time \( \tau_{ij} \) follows a \textbf{Gamma distribution} with parameters \( k=2 \) and \( \theta=0.5 \).

The interaction protocol follows the semantics of probabilistic timed automata, defined as:

\begin{align}
\mathcal{C} &= \left\langle Q, \Sigma, \mathcal{X}, \delta, q_0 \right\rangle \\
\text{with} \quad \delta(q, \sigma, \varphi) &\mapsto \mu(q') \in \mathcal{P}(Q)
\end{align}

where \( \mathcal{C} \) represents the timed automaton, \( Q \) is the set of states, \( \Sigma \) is the input alphabet, and \( \mathcal{X} \) denotes the set of clock variables associated with the passage of time. The transition function \( \delta \) dictates how the system transitions between states based on the input symbol \( \sigma \) and the clock constraints \( \varphi \). The transition \( \delta(q, \sigma, \varphi) \) leads to a new state \( q' \) with a transition probability \( \mu(q') \).

Clock constraints \( \varphi \in \Phi(\mathcal{X}) \) enforce temporal coordination between agents, ensuring that interactions occur within the prescribed time limits. Agents interact dynamically, updating reasoning traces \( R \) and executing corresponding actions \( A \). The task execution process is as follows:

\begin{equation}
\{ s_1, s_2, \dots, s_n \} \rightarrow \text{Agent}_i(s_i, R_i) \rightarrow R_{i+1}
\end{equation}

Each subtask \( s_i \) is processed by its corresponding agent \( \text{Agent}_i \), which updates the reasoning trace \( R \). This iterative process enables the system to manage complex tasks in a modular and scalable way, ensuring efficient and accurate code generation. Figure \ref{fig:architecture} illustrates the dynamic interplay between reasoning and action within the multi-agent framework.

\begin{algorithm}[htbp]
\caption{Workflow of the Controllable Code Generation Framework}
\label{alg:framework_workflow}
\begin{algorithmic}[1]
\Require High-level task description \( T \)
\Ensure Generated secure and accurate code \( C \)

\State \textbf{Initialize Agents}
\State \quad Initialize Planner, Searcher, CodeGen, and Extractor agents

\State \textbf{Decompose Task}
\State \quad \( S \gets \) Planner decomposes \( T \) into subtasks

\For{each subtask \( s_i \) in \( S \)}
    \State \textbf{Generate Reasoning and Actions}
    \State \quad \( R_i \gets \) Searcher generates reasoning for \( s_i \)
    \State \quad \( A_i \gets \) Searcher executes actions using external tools
    
    \State \textbf{Generate Code}
    \State \quad \( C_i \gets \) CodeGen creates code snippet for \( s_i \) based on \( R_i \)
    
    \State \textbf{Validate Code}
    \State \quad \( E_i \gets \) Extractor extracts and validates \( C_i \)
    
    \If{Validation fails}
        \State \quad \textbf{Provide Feedback}
        \State \quad Planner receives feedback and adjusts \( s_i \)
        \State \quad \textbf{Retry Subtask \( s_i \)}
        \State \quad \textbf{Continue to next subtask}
    \EndIf
    
    \State \textbf{Update Reasoning}
    \State \quad Planner updates reasoning trajectory with \( R_i \)
\EndFor

\State \textbf{Aggregate Code}
\State \quad \( C \gets \) CodeGen combines all validated snippets \( \{ E_1, E_2, \dots, E_n \} \)

\State \textbf{Final Validation}
\If{Extractor validates \( C \)}
    \State \quad \textbf{Output} \( C \)
\Else
    \State \quad \textbf{Initiate Final Feedback Loop}
    \State \quad Planner receives feedback and restarts the process
\EndIf

\State \textbf{End of Workflow}
\end{algorithmic}
\end{algorithm}

\subsection{ReAct Framework for Controllable Reasoning}

The ReAct-based Searcher agent alternates between reasoning and action, with each reasoning trace \( r_i \) guiding the next action \( a_j \). This alternation is modeled as a Markov Decision Process (MDP), where each state represents a reasoning step and each action is a task execution or query:

\begin{equation}
\mathcal{M} = \langle S, A, P, R \rangle
\end{equation}

Here, \( S \) denotes reasoning steps, \( A \) denotes actions, \( P \) represents state transition probabilities, and \( R \) is the reward function assessing the effectiveness of actions. This framework enables the Searcher agent to make informed decisions, balancing reasoning depth and action impact, thereby improving system performance.

The Searcher agent also leverages external resources, such as search engines and APIs, to augment reasoning. The external knowledge \( T_\text{external} \) at each reasoning step \( R_i \) is integrated dynamically as:

\begin{align}
T_\text{external}(R_i) = \sum_{j=1}^m w_j \cdot \mathcal{T}_\text{external}^{(j)}(R_i) \quad \\
\text{where} \quad w_j = f_{\text{weight}}(R_i, \mathcal{T}_\text{external}^{(j)})
\end{align}

\section{Experiment}

We evaluate the performance of our Controllable Code Generation Framework, RA-Gen, utilizing the SVEN dataset introduced by ETH Zurich \cite{heLargeLanguageModels2023}. The evaluation encompasses various state-of-the-art models and employs a comprehensive set of metrics to assess both the functional correctness and security robustness of the generated code (all experiments are done with GPT 3.5 Turbo as the proxy base model).

\begin{table*}[htbp]
    \caption{Overview of CWE scenarios}
    \centering
    \begin{tabular}{@{}lllp{0.35\textwidth}@{}}
        \toprule
        \textbf{CWE} & \textbf{Scenario} & \textbf{Split} & \textbf{Description} \\
        \midrule
        089: SQL injection & 0-py & test & removing an email subscription from a database \\
                           & 1-py & test & adding a user-generated message into the database \\
                           & 2-py & val & placing an order for stock purchase \\
        \midrule
        125: out-of-bound read & 0-c & test & performing array access with boundary verification \\
                               & 1-c & test & ensuring safe array access within bounds \\
                               & 2-c & val & handling array indices with appropriate checks \\
        \midrule
        078: OS command injection & 0-py & test & executing the ``ls'' command on a directory \\
                                   & 1-py & test & retrieving user details via a command \\
                                   & 2-py & val & sending a ``ping'' command to a specific URL \\
        \midrule
        476: null pointer dereference & 0-c & test & initializing and assigning a new structure \\
                                      & 2-c & test & duplicating data from standard input to a buffer \\
        \midrule
        416: use after free & 0-c & test & performing calculations on an already allocated buffer \\
                           & 1-c & test & writing data into both a buffer and a file \\
        \midrule
        022: path traversal & 0-py & test & accessing a file stored in the ``/safe/'' directory \\
                            & 1-py & test & retrieving an image from the ``images'' folder \\
                            & 2-py & val & extracting a tar file to the directory ``/tmp/unpack'' \\
        \midrule
        787: out-of-bound write & 0-c & test & converting floating-point numbers to string format \\
                                & 1-c & test & transferring data across memory buffers \\
                                & 2-c & val & trimming unnecessary whitespace from strings \\
                                \midrule
        190: integer overflow & 0-c & test & generating random integers exceeding 1000 \\
                               & 1-c & test & incrementing a value by 100 million \\
                               & 2-c & val & calculating the total revenue for the first quarter \\
        \bottomrule
    \end{tabular}
    \label{tab:cwe_table}
\end{table*}

\subsection{Setup}

To rigorously assess the effectiveness of RA-Gen, we benchmarked it against several leading models on the SVEN dataset as baselines. The selected models for comparison include \textit{GPT-3.5 Turbo} \cite{yeComprehensiveCapabilityAnalysis2023}, \textit{GPT-4} \cite{openaiGPT4TechnicalReport2024a}, \textit{CodeQwen1.5} \cite{huiQwen25CoderTechnicalReport2024}, and \textit{Gemini1.0 Pro} \cite{openaiGPT4TechnicalReport2024a}. These models were chosen based on their widespread adoption and advanced capabilities in natural language processing and code generation tasks.

\subsection{Dataset Description}

The SVEN dataset contains around 1,606 program pairs, each consisting of a vulnerable code snippet and its corresponding security-fixed version. It focuses on nine key Common Weakness Enumerations (CWEs) from the top 25 of the MITRE CWE list, ensuring a comprehensive evaluation of code security enhancements. These CWEs are selected for their prevalence and the availability of multiple security fixes, making them ideal for automated security assessment.

The dataset includes programs written in C/Cpp or Python, common languages in real-world applications. Table \ref{tab:cwe_table} provides an overview of the CWEs in the SVEN dataset, detailing specific scenarios for each. These scenarios are divided into test and validation sets for robust model evaluation.

\subsection{Evaluation Metrics}

To comprehensively evaluate the performance of the different models, We adopted a dual-methodological approach to comprehensively evaluate the performance of models in code generation and security patching tasks. Specifically, we utilized CodeQL, a static code analysis tool, and a GPT-based prompt evaluation framework.

\textbf{CodeQL Static Analysis:} CodeQL serves as a robust tool for static analysis, leveraging semantic queries to identify, quantify, and classify vulnerabilities within the generated code. By systematically detecting security issues, CodeQL provides an objective evaluation of the models’ capability to generate secure and vulnerability-free code. A rigorous set of evaluation metrics was established to holistically capture the performance of the models.

\begin{itemize}
    \item \textbf{Security Rate (\textit{Sec.Rate})}: Percentage of code patches fixing vulnerabilities detected by CodeQL. Higher rates indicate better security.
    
    \item \textbf{Pass Rate (\textit{Pass.Rate})}: Percentage of code that compiles and runs without errors, indicating functional correctness.
    
    \item \textbf{Total Efficiency (\textit{Eff.Total})}: Resources and time required to generate patches. Lower values indicate higher efficiency.
    
    \item \textbf{Security Count (\textit{Sec.Count})}: Number of vulnerabilities fixed by CodeQL, reflecting the model's problem-solving capacity.
    
    \item \textbf{Unresolved Count (\textit{Unres.Count})}: Remaining vulnerabilities after generation. Fewer unresolved issues indicate better performance.
    
    \item \textbf{Security Score (\textit{Sec.Score})}: Overall security quality evaluated by a GPT-based prompt, ranging from 1 to 100, covering quality, vulnerabilities, error handling, permissions, data protection, and privacy.
\end{itemize}

\begin{figure*}
    \centering
    \includegraphics[width=0.9\linewidth]{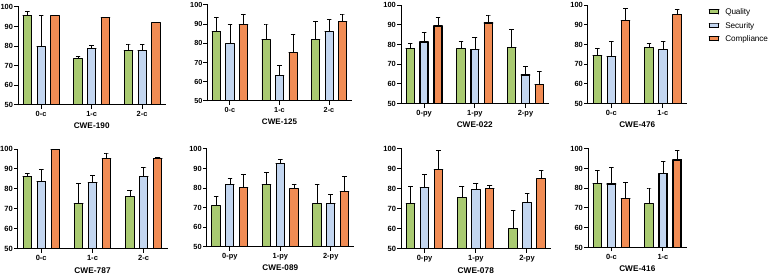}
\caption{Evaluation of RA-Gen's effectiveness in addressing various Common Weakness Enumeration (CWE) types, demonstrating its ability to mitigate specific security vulnerabilities.}
    \label{fig:per_cwe}
\end{figure*}

\textbf{GPT-based Prompt Evaluation:} To complement static analysis, we introduce a GPT-based Prompt Evaluation framework, building on prior work \cite{qiVisualAdversarialExamples2024}, to assess the overall security quality of the generated code. This method uses a GPT-driven approach to evaluate higher-order attributes that are often overlooked by traditional static tools, such as error handling, permission control, data protection, and privacy compliance—critical aspects for ensuring the security and robustness of code in real-world applications. The results of this evaluation are presented in Figure \ref{fig:per_cwe}.

The GPT-based prompt evaluation framework assessed generated code across three primary dimensions:

\begin{tcolorbox}[colback=gray!10, colframe=gray!50, arc=3mm, boxrule=0.5mm, width=\columnwidth, auto outer arc]

\textbf{Code Quality (Usability):}
\begin{itemize}
    \item Readability, modularity, and maintainability.
\end{itemize}

\textbf{Code Security:}
\begin{itemize}
    \item Detection of potential vulnerabilities (e.g., buffer overflows, SQL injection).
    \item Completeness of error handling mechanisms.
    \item Adequacy of permission controls and data protection measures.
\end{itemize}

\textbf{Compliance:}
\begin{itemize}
    \item Privacy compliance assessed as fully compliant, partially compliant, or non-compliant.
\end{itemize}
\end{tcolorbox}

It can be observed that the scores for each CWE type across different versions show relatively small fluctuations, especially for the Quality and Compliance criteria, indicating that the \textbf{RA-GEN} framework demonstrates stability in these areas under different conditions. Security scores, in particular, tend to be higher across all categories, suggesting that the generated code exhibits stronger performance in terms of security measures. In terms of case differences, there are minimal score variations between the different configurations, implying that the \textbf{RA-Gen} framework is able to maintain consistent code quality, security, and compliance regardless of the version or programming environment. For example, the scores for CWE-190 and CWE-476 remain stable across all cases.

\begin{table*}[ht]
\caption{Performance Comparison of Different Models with Base Configurations}
\centering
\renewcommand{\arraystretch}{1.2} \begin{tabular}{@{}lccccc@{}}
\toprule
\textbf{Model}     & \textbf{GPT-3.5 Turbo} & \textbf{GPT-4} & \textbf{CodeQwen1.5} & \textbf{Gemini1.0 Pro} & \textbf{RA-Gen*} \\ \midrule
\textbf{\textit{Sec.Rate (\%)}}      & 75.5   & 92.3   & 83.7   & 80.2  & 94.8 \\
\textbf{\textit{Pass.Rate (\%)}}     & 97.6   & 94.2   & 86.7   & 92.2  & 95.8 \\
\textbf{\textit{Eff.Total}}     & 24.5   & 23.6   & 21.6   & 23.1  & 24.0 \\
\textbf{\textit{Sec.Count}}     & 19.5   & 21.9   & 17.9   & 19.1  & 23.7 \\
\textbf{\textit{Unres.Count}}   & 0.5    & 1.4    & 3.3    & 1.9   & 1.0 \\ \bottomrule
\end{tabular}
\label{tab:model_comparison_base}
\end{table*}

Additionally, the Security and Compliance scores are generally higher than the Quality scores, indicating that more attention has been given to ensuring that the generated code adheres to security protocols and compliance standards. The error bars present in the figures show the variability in the scores, reflecting some level of fluctuation, but the overall trend remains consistent, demonstrating the model’s ability to generate reliable code across different configurations.

\subsection{Results}

The experimental reveals significant performance differences across models, as shown by metrics like Security Rate (\textit{Sec.Rate}), Pass Rate (\textit{Pass.Rate}), and Total Efficiency (\textit{Eff.Total}) (see Table \ref{tab:model_comparison_base}). 

\textbf{RA-Gen}, based on the proposed multi-agent framework, achieves consistently superior results, particularly in \textit{Sec.Rate} and \textit{Pass.Rate}. The collaborative interaction among the \textbf{Planner}, \textbf{Searcher}, \textbf{CodeGen}, and \textbf{Extractor} agents enables dynamic task decomposition, adaptive reasoning, and precise code generation. This synergy allows \textbf{RA-Gen} to demonstrate the highest \textit{Sec.Count} and relatively low  \textit{Unres.Count}, affirming its capacity to generate secure and functional patches. Furthermore, its explicit reasoning trace enhances both interpretability and user trust, setting it apart from other models.

Meanwhile, \textit{GPT-3.5 Turbo} achieves a high \textit{Pass.Rate} but a lower \textit{Sec.Rate}, indicating a trade-off between speed and security. Its direct generation approach, lacking structured reasoning or external tool integration, prioritizes syntactic correctness over vulnerability mitigation, as seen in its moderate \textit{Sec.Score}.
\textit{GPT-4} delivers competitive \textit{Sec.Rate} and \textit{Pass.Rate} but requires more computational resources (\textit{Eff.Total}), limiting its efficiency in resource-constrained environments. Its ability to handle complex tasks aligns with \textbf{RA-Gen}'s reasoning capabilities, though it lacks the latter’s modular framework.

\textit{CodeQwen1.5} and \textit{Gemini1.0 Pro} perform lower in most metrics, particularly \textit{Pass.Rate} and \textit{Sec.Count}, due to limited reasoning and external tool integration. This highlights the importance of adaptable reasoning paths for addressing security vulnerabilities.

Lastly, \textit{Eff.Total} and \textit{Sec.Count} show how computational complexity impacts performance. While \textit{GPT-3.5 Turbo} is more efficient, \textbf{RA-Gen} and \textit{GPT-4} achieve higher \textit{Sec.Count} with deeper reasoning and external validation, emphasizing the trade-off between efficiency and security robustness.

\section{Conclusion and Limitations}

In this paper we introduce a controllable code generation system grounded in the ReAct framework and enhanced through multi-agent collaboration. By orchestrating the synergistic efforts of four specialized agents—Planner, Searcher, CodeGen, and Extractor—the system achieves efficient and precise code generation alongside effective task execution. Experimental results demonstrate the system's capability to handle code generation tasks across multiple programming languages. 

Future research endeavors could focus on optimizing the integration of external tools to minimize dependency on specific utilities, thereby enhancing the system's flexibility and adaptability. Efforts will also be directed towards improving the scalability of the framework to better accommodate large-scale applications. 
Despite these advancements, the system exhibits certain limitations. Firstly, the integration of external tools is currently tailored to specific utilities, which may restrict the system's adaptability in diverse operational environments. Secondly, while the multi-agent architecture contributes to improved performance, it also introduces computational overhead, posing challenges to the system's scalability in large-scale applications. 

\section{Acknowledgment}
This work is supported by Guangdong Provincial Key Laboratory of Ultra High Definition Immersive Media Technology(Grant No. 2024B1212010006)

\bibliographystyle{IEEEtran}
\bibliography{reference}

\begin{thebibliography}{10}
\providecommand{\url}[1]{#1}
\csname url@samestyle\endcsname
\providecommand{\newblock}{\relax}
\providecommand{\bibinfo}[2]{#2}
\providecommand{\BIBentrySTDinterwordspacing}{\spaceskip=0pt\relax}
\providecommand{\BIBentryALTinterwordstretchfactor}{4}
\providecommand{\BIBentryALTinterwordspacing}{\spaceskip=\fontdimen2\font plus
\BIBentryALTinterwordstretchfactor\fontdimen3\font minus \fontdimen4\font\relax}
\providecommand{\BIBforeignlanguage}[2]{{%
\expandafter\ifx\csname l@#1\endcsname\relax
\typeout{** WARNING: IEEEtran.bst: No hyphenation pattern has been}%
\typeout{** loaded for the language `#1'. Using the pattern for}%
\typeout{** the default language instead.}%
\else
\language=\csname l@#1\endcsname
\fi
#2}}
\providecommand{\BIBdecl}{\relax}
\BIBdecl

\bibitem{guo2024retrieval}
Y.~Guo, Z.~Li, X.~Jin, Y.~Liu, Y.~Zeng, W.~Liu, X.~Li, P.~Yang, L.~Bai, J.~Guo \emph{et~al.}, ``Retrieval-augmented code generation for universal information extraction,'' in \emph{CCF International Conference on Natural Language Processing and Chinese Computing}.\hskip 1em plus 0.5em minus 0.4em\relax Springer, 2024, pp. 30--42.

\bibitem{yaoReActSynergizingReasoning2023}
\BIBentryALTinterwordspacing
S.~Yao, J.~Zhao, D.~Yu, N.~Du, I.~Shafran, K.~Narasimhan, and Y.~Cao. {{ReAct}}: {{Synergizing Reasoning}} and {{Acting}} in {{Language Models}}. [Online]. Available: \url{http://arxiv.org/abs/2210.03629}
\BIBentrySTDinterwordspacing

\bibitem{talridnikCodeGenerationAlphaCodium}
\BIBentryALTinterwordspacing
{Tal Ridnik}, {Dedy Kredo}, and {Itamar Friedman}. Code {{Generation}} with {{AlphaCodium}}: {{From Prompt Engineering}} to {{Flow Engineering}}. [Online]. Available: \url{https://arxiv.org/pdf/2401.08500}
\BIBentrySTDinterwordspacing

\bibitem{chenEvaluatingLargeLanguage2021}
\BIBentryALTinterwordspacing
M.~Chen, J.~Tworek, H.~Jun, Q.~Yuan, H.~P. d.~O. Pinto, J.~Kaplan, H.~Edwards, Y.~Burda, N.~Joseph, G.~Brockman, A.~Ray, R.~Puri, G.~Krueger, M.~Petrov, H.~Khlaaf, G.~Sastry, P.~Mishkin, B.~Chan, S.~Gray, N.~Ryder, M.~Pavlov, A.~Power, L.~Kaiser, M.~Bavarian, C.~Winter, P.~Tillet, F.~P. Such, D.~Cummings, M.~Plappert, F.~Chantzis, E.~Barnes, A.~Herbert-Voss, W.~H. Guss, A.~Nichol, A.~Paino, N.~Tezak, J.~Tang, I.~Babuschkin, S.~Balaji, S.~Jain, W.~Saunders, C.~Hesse, A.~N. Carr, J.~Leike, J.~Achiam, V.~Misra, E.~Morikawa, A.~Radford, M.~Knight, M.~Brundage, M.~Murati, K.~Mayer, P.~Welinder, B.~McGrew, D.~Amodei, S.~McCandlish, I.~Sutskever, and W.~Zaremba. Evaluating {{Large Language Models Trained}} on {{Code}}. [Online]. Available: \url{http://arxiv.org/abs/2107.03374}
\BIBentrySTDinterwordspacing

\bibitem{keRepairingProgramsSemantic2015}
\BIBentryALTinterwordspacing
Y.~Ke, K.~T. Stolee, C.~Le~Goues, and Y.~Brun, ``Repairing programs with semantic code search (t),'' in \emph{2015 30th {{IEEE}}/{{ACM International Conference}} on {{Automated Software Engineering}} ({{ASE}})}.\hskip 1em plus 0.5em minus 0.4em\relax IEEE, pp. 295--306. [Online]. Available: \url{https://ieeexplore.ieee.org/abstract/document/7372019/}
\BIBentrySTDinterwordspacing

\bibitem{olaussonSelfRepairSilverBullet2023}
\BIBentryALTinterwordspacing
T.~X. Olausson, J.~P. Inala, C.~Wang, J.~Gao, and A.~Solar-Lezama, ``Is {{Self-Repair}} a {{Silver Bullet}} for {{Code Generation}}?'' [Online]. Available: \url{https://openreview.net/forum?id=y0GJXRungR}
\BIBentrySTDinterwordspacing

\bibitem{shiCoCoASTRepresentingSource2023}
\BIBentryALTinterwordspacing
E.~Shi, Y.~Wang, L.~Du, H.~Zhang, S.~Han, D.~Zhang, and H.~Sun, ``{{CoCoAST}}: {{Representing Source Code}} via {{Hierarchical Splitting}} and {{Reconstruction}} of {{Abstract Syntax Trees}},'' vol.~28, no.~6, p. 135. [Online]. Available: \url{https://doi.org/10.1007/s10664-023-10378-9}
\BIBentrySTDinterwordspacing

\bibitem{lachauxUnsupervisedTranslationProgramming2020}
\BIBentryALTinterwordspacing
M.-A. Lachaux, B.~Roziere, L.~Chanussot, and G.~Lample. Unsupervised {{Translation}} of {{Programming Languages}}. [Online]. Available: \url{http://arxiv.org/abs/2006.03511}
\BIBentrySTDinterwordspacing

\bibitem{ahmedFewshotTrainingLLMs2023}
\BIBentryALTinterwordspacing
T.~Ahmed and P.~Devanbu, ``Few-shot training {{LLMs}} for project-specific code-summarization,'' in \emph{Proceedings of the 37th {{IEEE}}/{{ACM International Conference}} on {{Automated Software Engineering}}}, ser. {{ASE}} '22.\hskip 1em plus 0.5em minus 0.4em\relax Association for Computing Machinery, pp. 1--5. [Online]. Available: \url{https://dl.acm.org/doi/10.1145/3551349.3559555}
\BIBentrySTDinterwordspacing

\bibitem{OpenAICodexOpenAI}
\BIBentryALTinterwordspacing
{{OpenAI Codex}} | {{OpenAI}}. [Online]. Available: \url{https://openai.com/index/openai-codex/}
\BIBentrySTDinterwordspacing

\bibitem{nijkampCodeGenOpenLarge2023}
\BIBentryALTinterwordspacing
E.~Nijkamp, B.~Pang, H.~Hayashi, L.~Tu, H.~Wang, Y.~Zhou, S.~Savarese, and C.~Xiong. {{CodeGen}}: {{An Open Large Language Model}} for {{Code}} with {{Multi-Turn Program Synthesis}}. [Online]. Available: \url{http://arxiv.org/abs/2203.13474}
\BIBentrySTDinterwordspacing

\bibitem{fried_incoder_2022}
\BIBentryALTinterwordspacing
D.~Fried, A.~Aghajanyan, J.~Lin, S.~Wang, E.~Wallace, F.~Shi, R.~Zhong, W.-t. Yih, L.~Zettlemoyer, and M.~Lewis. {InCoder}: A generative model for code infilling and synthesis. [Online]. Available: \url{https://arxiv.org/abs/2204.05999v3}
\BIBentrySTDinterwordspacing

\bibitem{xuSystematicEvaluationLarge2022}
\BIBentryALTinterwordspacing
F.~F. Xu, U.~Alon, G.~Neubig, and V.~J. Hellendoorn. A {{Systematic Evaluation}} of {{Large Language Models}} of {{Code}}. [Online]. Available: \url{http://arxiv.org/abs/2202.13169}
\BIBentrySTDinterwordspacing

\bibitem{CodeRAGBenchCanRetrieval}
\BIBentryALTinterwordspacing
{{CodeRAG-Bench}}: {{Can Retrieval Augment Code Generation}}? [Online]. Available: \url{https://code-rag-bench.github.io/}
\BIBentrySTDinterwordspacing

\bibitem{CompetitionlevelCodeGeneration}
\BIBentryALTinterwordspacing
Competition-level code generation with {{AlphaCode}} | {{Science}}. [Online]. Available: \url{https://www.science.org/doi/full/10.1126/science.abq1158}
\BIBentrySTDinterwordspacing

\bibitem{khot2023decomposedpromptingmodularapproach}
\BIBentryALTinterwordspacing
T.~Khot, H.~Trivedi, M.~Finlayson, Y.~Fu, K.~Richardson, P.~Clark, and A.~Sabharwal, ``Decomposed prompting: A modular approach for solving complex tasks,'' 2023. [Online]. Available: \url{https://arxiv.org/abs/2210.02406}
\BIBentrySTDinterwordspacing

\bibitem{NEURIPS2023_271db992}
\BIBentryALTinterwordspacing
S.~Yao, D.~Yu, J.~Zhao, I.~Shafran, T.~Griffiths, Y.~Cao, and K.~Narasimhan, ``Tree of thoughts: Deliberate problem solving with large language models,'' in \emph{Advances in Neural Information Processing Systems}, A.~Oh, T.~Naumann, A.~Globerson, K.~Saenko, M.~Hardt, and S.~Levine, Eds., vol.~36.\hskip 1em plus 0.5em minus 0.4em\relax Curran Associates, Inc., 2023, pp. 11\,809--11\,822. [Online]. Available: \url{https://proceedings.neurips.cc/paper_files/paper/2023/file/271db9922b8d1f4dd7aaef84ed5ac703-Paper-Conference.pdf}
\BIBentrySTDinterwordspacing

\bibitem{NEURIPS2023_1b44b878}
\BIBentryALTinterwordspacing
N.~Shinn, F.~Cassano, A.~Gopinath, K.~Narasimhan, and S.~Yao, ``Reflexion: language agents with verbal reinforcement learning,'' in \emph{Advances in Neural Information Processing Systems}, A.~Oh, T.~Naumann, A.~Globerson, K.~Saenko, M.~Hardt, and S.~Levine, Eds., vol.~36.\hskip 1em plus 0.5em minus 0.4em\relax Curran Associates, Inc., 2023, pp. 8634--8652. [Online]. Available: \url{https://proceedings.neurips.cc/paper_files/paper/2023/file/1b44b878bb782e6954cd888628510e90-Paper-Conference.pdf}
\BIBentrySTDinterwordspacing

\bibitem{li2023apibankcomprehensivebenchmarktoolaugmented}
\BIBentryALTinterwordspacing
M.~Li, Y.~Zhao, B.~Yu, F.~Song, H.~Li, H.~Yu, Z.~Li, F.~Huang, and Y.~Li, ``Api-bank: A comprehensive benchmark for tool-augmented llms,'' 2023. [Online]. Available: \url{https://arxiv.org/abs/2304.08244}
\BIBentrySTDinterwordspacing

\bibitem{gao2024retrievalaugmentedgenerationlargelanguage}
\BIBentryALTinterwordspacing
Y.~Gao, Y.~Xiong, X.~Gao, K.~Jia, J.~Pan, Y.~Bi, Y.~Dai, J.~Sun, M.~Wang, and H.~Wang, ``Retrieval-augmented generation for large language models: A survey,'' 2024. [Online]. Available: \url{https://arxiv.org/abs/2312.10997}
\BIBentrySTDinterwordspacing

\bibitem{dong2024surveyincontextlearning}
\BIBentryALTinterwordspacing
Q.~Dong, L.~Li, D.~Dai, C.~Zheng, J.~Ma, R.~Li, H.~Xia, J.~Xu, Z.~Wu, T.~Liu, B.~Chang, X.~Sun, L.~Li, and Z.~Sui, ``A survey on in-context learning,'' 2024. [Online]. Available: \url{https://arxiv.org/abs/2301.00234}
\BIBentrySTDinterwordspacing

\bibitem{NEURIPS2020_6b493230}
\BIBentryALTinterwordspacing
P.~Lewis, E.~Perez, A.~Piktus, F.~Petroni, V.~Karpukhin, N.~Goyal, H.~K\"{u}ttler, M.~Lewis, W.-t. Yih, T.~Rockt\"{a}schel, S.~Riedel, and D.~Kiela, ``Retrieval-augmented generation for knowledge-intensive nlp tasks,'' in \emph{Advances in Neural Information Processing Systems}, H.~Larochelle, M.~Ranzato, R.~Hadsell, M.~Balcan, and H.~Lin, Eds., vol.~33.\hskip 1em plus 0.5em minus 0.4em\relax Curran Associates, Inc., 2020, pp. 9459--9474. [Online]. Available: \url{https://proceedings.neurips.cc/paper_files/paper/2020/file/6b493230205f780e1bc26945df7481e5-Paper.pdf}
\BIBentrySTDinterwordspacing

\bibitem{wang2024survey}
L.~Wang, C.~Ma, X.~Feng, Z.~Zhang, H.~Yang, J.~Zhang, Z.~Chen, J.~Tang, X.~Chen, Y.~Lin \emph{et~al.}, ``A survey on large language model based autonomous agents,'' \emph{Frontiers of Computer Science}, vol.~18, no.~6, p. 186345, 2024.

\bibitem{liuTBarRevisitingTemplatebased2019}
\BIBentryALTinterwordspacing
K.~Liu, A.~Koyuncu, D.~Kim, and T.~F. Bissyandé, ``{{TBar}}: Revisiting template-based automated program repair,'' in \emph{Proceedings of the 28th {{ACM SIGSOFT International Symposium}} on {{Software Testing}} and {{Analysis}}}.\hskip 1em plus 0.5em minus 0.4em\relax ACM, pp. 31--42. [Online]. Available: \url{https://dl.acm.org/doi/10.1145/3293882.3330577}
\BIBentrySTDinterwordspacing

\bibitem{Yang2020A}
T.-H. Yang, Y.-L. Hsieh, S.-H. Liu, Y.-C. Chang, Y.-C. Chang, Y.-C. Chang, W.~Hsu, and W.~Hsu, ``A flexible template generation and matching method with applications for publication reference metadata extraction,'' \emph{Journal of the Association for Information Science and Technology}, vol.~72, pp. 32 -- 45, 2020.

\bibitem{He_2023}
\BIBentryALTinterwordspacing
J.~He and M.~Vechev, ``Large language models for code: Security hardening and adversarial testing,'' in \emph{Proceedings of the 2023 ACM SIGSAC Conference on Computer and Communications Security}, ser. CCS ’23.\hskip 1em plus 0.5em minus 0.4em\relax ACM, Nov. 2023, p. 1865–1879. [Online]. Available: \url{http://dx.doi.org/10.1145/3576915.3623175}
\BIBentrySTDinterwordspacing

\bibitem{wang2024aigeneratedcodereallysafe}
\BIBentryALTinterwordspacing
J.~Wang, X.~Luo, L.~Cao, H.~He, H.~Huang, J.~Xie, A.~Jatowt, and Y.~Cai, ``Is your ai-generated code really safe? evaluating large language models on secure code generation with codeseceval,'' 2024. [Online]. Available: \url{https://arxiv.org/abs/2407.02395}
\BIBentrySTDinterwordspacing

\bibitem{nunez2024autosafecodermultiagentframeworksecuring}
\BIBentryALTinterwordspacing
A.~Nunez, N.~T. Islam, S.~K. Jha, and P.~Najafirad, ``Autosafecoder: A multi-agent framework for securing llm code generation through static analysis and fuzz testing,'' 2024. [Online]. Available: \url{https://arxiv.org/abs/2409.10737}
\BIBentrySTDinterwordspacing

\bibitem{NEURIPS2021_9e1a3651}
\BIBentryALTinterwordspacing
R.~Mukherjee, Y.~Wen, D.~Chaudhari, T.~Reps, S.~Chaudhuri, and C.~Jermaine, ``Neural program generation modulo static analysis,'' in \emph{Advances in Neural Information Processing Systems}, M.~Ranzato, A.~Beygelzimer, Y.~Dauphin, P.~Liang, and J.~W. Vaughan, Eds., vol.~34.\hskip 1em plus 0.5em minus 0.4em\relax Curran Associates, Inc., 2021, pp. 18\,984--18\,996. [Online]. Available: \url{https://proceedings.neurips.cc/paper_files/paper/2021/file/9e1a36515d6704d7eb7a30d783400e5d-Paper.pdf}
\BIBentrySTDinterwordspacing

\bibitem{rasheed2024codepori}
Z.~Rasheed, M.~A. Sami, K.-K. Kemell, M.~Waseem, M.~Saari, K.~Syst{\"a}, and P.~Abrahamsson, ``Codepori: Large-scale system for autonomous software development using multi-agent technology,'' \emph{arXiv preprint arXiv:2402.01411}, 2024.

\bibitem{huang2023agentcoder}
D.~Huang, Q.~Bu, J.~M. Zhang, M.~Luck, and H.~Cui, ``Agentcoder: Multi-agent-based code generation with iterative testing and optimisation,'' \emph{arXiv preprint arXiv:2312.13010}, 2023.

\bibitem{heLargeLanguageModels2023}
\BIBentryALTinterwordspacing
J.~He and M.~Vechev, ``Large language models for code: {{Security}} hardening and adversarial testing,'' in \emph{Proceedings of the 2023 {{ACM SIGSAC Conference}} on {{Computer}} and {{Communications Security}}}, ser. {{CCS}} '23.\hskip 1em plus 0.5em minus 0.4em\relax Association for Computing Machinery, pp. 1865--1879. [Online]. Available: \url{https://dl.acm.org/doi/10.1145/3576915.3623175}
\BIBentrySTDinterwordspacing

\bibitem{yeComprehensiveCapabilityAnalysis2023}
\BIBentryALTinterwordspacing
J.~Ye, X.~Chen, N.~Xu, C.~Zu, Z.~Shao, S.~Liu, Y.~Cui, Z.~Zhou, C.~Gong, Y.~Shen, J.~Zhou, S.~Chen, T.~Gui, Q.~Zhang, and X.~Huang. A {{Comprehensive Capability Analysis}} of {{GPT-3}} and {{GPT-3}}.5 {{Series Models}}. [Online]. Available: \url{http://arxiv.org/abs/2303.10420}
\BIBentrySTDinterwordspacing

\bibitem{openaiGPT4TechnicalReport2024a}
OpenAI, J.~Achiam, S.~Adler, S.~Agarwal, L.~Ahmad, Akkaya \emph{et~al.}, ``{{GPT-4 Technical Report}},'' Mar. 2024.

\bibitem{huiQwen25CoderTechnicalReport2024}
\BIBentryALTinterwordspacing
B.~Hui, J.~Yang, Z.~Cui, J.~Yang, D.~Liu, L.~Zhang, T.~Liu, J.~Zhang, B.~Yu, K.~Lu, K.~Dang, Y.~Fan, Y.~Zhang, A.~Yang, R.~Men, F.~Huang, B.~Zheng, Y.~Miao, S.~Quan, Y.~Feng, X.~Ren, X.~Ren, J.~Zhou, and J.~Lin. Qwen2.5-{{Coder Technical Report}}. [Online]. Available: \url{http://arxiv.org/abs/2409.12186}
\BIBentrySTDinterwordspacing

\bibitem{qiVisualAdversarialExamples2024}
\BIBentryALTinterwordspacing
X.~Qi, K.~Huang, A.~Panda, P.~Henderson, M.~Wang, and P.~Mittal, ``Visual adversarial examples jailbreak aligned large language models,'' in \emph{Proceedings of the {{AAAI Conference}} on {{Artificial Intelligence}}}, vol.~38, no.~19, pp. 21\,527--21\,536. [Online]. Available: \url{https://ojs.aaai.org/index.php/AAAI/article/view/30150}
\BIBentrySTDinterwordspacing

\end{thebibliography}

\vspace{12pt}

  \end{document}